\documentclass[prl,notitlepage,twocolumn,groupedaddress]{revtex4-1}

\usepackage{epsfig,color}
\usepackage{graphicx}
\usepackage{dcolumn}% Align table columns on decimal point
\usepackage{bm}% bold math
\usepackage{amsmath, amsfonts, amssymb,mathrsfs}
\usepackage{pstricks}
\usepackage{amsxtra}
\usepackage{amsthm}
\usepackage{natbib}
\usepackage{qcircuit}
\usepackage{array}

\def\be{\begin{equation}}      % with numbering
\def\ee{\end{equation}}
\def\beu{\begin{equation*}}   % without numbering
\def\eeu{\end{equation*}}

\providecommand{\abs}[1]{\left\lvert#1\right\rvert}   % absolute value
\DeclareMathOperator{\trace}{Tr}      % Trace of matrix
\providecommand{\mean}[1]{\langle#1\rangle}
\providecommand{\bx}{{\bm{x}}}
\providecommand{\by}{{\bm{y}}}

\theoremstyle{definition}

\definecolor{new}{rgb}{.08,.05,.8}

\newcommand{\delete}[1]{{}}

\begin{document}
\title{Localization as an entanglement phase transition in boundary-driven Anderson models}
\date{\today}
\author{Michael J. Gullans}
\author{David A. Huse}
\affiliation{Department of Physics, Princeton University, Princeton, New Jersey 08544, USA}

\begin{abstract}
The Anderson localization transition is one of the most well studied examples of a zero temperature quantum phase transition.  On the other hand, many open questions remain about the phenomenology of disordered systems  driven far out of equilibrium.  Here we study the localization transition in the prototypical three-dimensional, noninteracting Anderson model when the system is driven at its boundaries to induce a current carrying non-equilibrium steady state.  Recently we showed that the diffusive phase of this model exhibits extensive mutual information of its non-equilibrium steady-state density matrix.  We show that that this extensive scaling persists in the entanglement and at the localization critical point, before crossing over to a short-range (area-law) scaling in the localized phase.  We introduce an entanglement witness for fermionic states that we name the mutual coherence, which, for fermionic Gaussian states, is also a lower bound on the mutual information.  Through a combination of analytical arguments and numerics, we  determine the finite-size scaling of the mutual coherence across the transition.  These results further develop the notion of entanglement phase transitions in open systems, with direct implications for driven many-body localized systems, as well as experimental studies of driven-disordered systems.
\end{abstract}
\maketitle

The notion that the entropy due to entanglement can be extensive in quantum many-body systems came into sharp focus with the introduction of the eigenstate thermalization hypothesis (ETH), which postulates that even single eigenstates of thermalizing (chaotic) Hamiltonians are in thermal equilibrium \cite{Deutsch91,Srednicki94,DAlessio16}.  Macroscopic thermodynamic entropy arises in this formulation through intrinsic extensive (``volume-law'') entanglement of the eigenstates.  Historically, these concepts arose from studying foundational questions in statistical mechanics and quantum aspects of black hole thermodynamics \cite{Bombelli86,Srednicki93}; however, advances in isolating and controlling quantum many-body systems  now allow these foundational concepts about the role of entanglement in statistical mechanics to be tested experimentally through both direct measurements \cite{Moura04,Daley12,Islam15,Pichler16,Kaufman16,Lukin18} and indirect methods \cite{Larkin65,Jurcevic14,Richerme14,Fukuhara15,Swingle16,Yao16,Zhu16,Li17,Wei18,Garttner17,Garttner18,Niknam18}.

However, there are also many situations where the entanglement entropy is not extensive.  This includes mixed-state density operators of thermal equilibrium Gibbs states and ground states of many local Hamiltonians \cite{Wolf08,Eisert10}, as well as eigenstates of systems that are many-body localized (MBL) \cite{Huse14,Serbyn13,Imbrie16,Imbrie16b}.  In the latter case, there is an {\it entanglement phase transition} at the MBL transition between extensive eigenstate entanglement in the ETH-obeying thermal phase and sub-extensive (only boundary-law) entanglement in the MBL phase where the ETH is violated \cite{Basko06,Gornyi05,Pal10,Nandkishore15}. Other examples of entanglement phase transitions have been analyzed in a random tensor network model \cite{Vasseur18} and in quantum circuit models with measurements \cite{Aharonov00,Li18,Skinner18,Chan18b}.   Due to the fundamental difficulty in distinguishing classical and quantum correlations in mixed states \cite{Horodecki09}, the entanglement properties of many-body mixed state density operators in microscopic models have generally been less studied than pure states, but there are examples of boundary-driven open systems with extensive entanglement in their long time states \cite{Gullans18}.

In this Letter, we further develop the phenomenology of entanglement phase transitions in open systems by studying the Anderson localization transition from this perspective.  We consider the prototypical case of single-particle Anderson localization on a three-dimensional lattice with a quenched random potential \cite{Anderson58}.  But we study this as a noninteracting many-fermion open system that is boundary-driven.  The driving is by clean conducting leads with incoming scattering states populated at different chemical potentials at the two ends of a disordered ``sample.''  We examine the non-equilibrium steady state (NESS) of this driven open system.  
%\new{As we detail below, our theoretical approach departs from many of the prior studies on Anderson localization transitions, while also providing novel insights into this broad class of disorder-driven quantum phase transitions.}

Recently, we showed that the NESS density matrix exhibits volume-law mutual information in the diffusive phase of this system \cite{Gullans18}.  Here, we extend this analysis to study the entanglement, as well as the localized phase and the localization critical point.  We find that the localized phase exhibits area-law mutual information, as might be expected.  We find that the mutual information remains volume-law at the critical point and in the diffusive phase.  Throughout this work, we use an entanglement witness for fermionic states that we introduce here and name the ``mutual coherence.''  For Gaussian fermionic states, as are encountered in the boundary-driven Anderson model, the mutual coherence is a lower bound on the mutual information.
 Combining single-parameter scaling theory and numerics with simple physical  arguments based on the production, spreading, and decoherence of operators in this system, we determine the finite-size critical scaling of the mutual coherence through the localization transition.  Due to the relative dearth of examples of non-equilibrium phase transitions where entanglement density serves as an order parameter, we believe this example can serve as a useful point of reference, with potentially immediate consequences for the analysis of current-driven MBL systems \cite{Znidaric16,Setiawan17,Varma17,Buca18}.  In addition, these results are broadly applicable to noninteracting models of disordered systems, making our predictions experimentally testable in a wide range of physical systems on mesoscopic length scales.

Although the Anderson model was originally introduced in 1958  \cite{Anderson58}, systematic investigations of  metal-insulator transitions in noninteracting versions of these models only began in the 1970's (for an overview see Ref. \cite{Evers08}).  Since that time, there has been continued progress on understanding these transitions from a variety of angles including approximate field theory descriptions \cite{EfetovBook}, numerical computations \cite{Markos06}, and rigorous mathematics \cite{Spencer10}.  Despite this sustained effort, the effects we describe in this work have, to our knowledge, not been previously identified.  We believe the reason for this omission is that the point of departure for our analysis is rather unconventional in that we are interested in the many-body state of the fermions when they are driven out of equilibrium by a chemical potential bias.   On the other hand, much of the literature on noninteracting Anderson models has focused on near equilibrium response functions, which can be characterized in terms of few-particle equilibrium Green's functions.  Spectral and spatial statistics (including entanglement properties \cite{Jia08}) of single-particle wavefunctions at criticality have been extensively analyzed \cite{Janssen94,Huckestein95}; however, the effects considered in this work only appear when performing weighted sums over all single-particle scattering states, with non-equilibrium populations.  Because part of the motivation for this work is to gain insights into noninteracting Anderson localization transitions that may also apply to interacting systems and MBL, we focus on arguments rooted in random quantum circuit models \cite{Tibor17,Khemani17,Gullans18}, which are more easily generalized to account for interactions \cite{Nahum16,Nahum17,vonKeyserlingk17,Nahum18b}.  In the supplemental material, we present an alternative derivation of the entanglement scaling analysis that more directly connects to past work on Anderson models \cite{supp}.  In both cases, we find that the volume-law mutual information and entanglement that builds up at the critical point and in the diffusive phase arises from a subtle interplay between the production and decoherence of long-range correlations, a key insight of our work, that has not been appreciated in past work on this problem.

Despite some similarities, there are a number of crucial distinctions between the entanglement phase transition studied in this work and the eigenstate entanglement transition studied in MBL.  One  difference is that here we consider the single mixed NESS of an open quantum system driven out of equilibrium, whereas the MBL transition occurs for exponentially many eigenstates of a closed quantum system.   A second important distinction is that the volume-law entanglement found here in the diffusive phase relies on the many-body system being noninteracting: according to our previous analysis, an interacting driven and diffusive system should have only area-law entanglement \cite{Gullans18}.  The phases in the thermal-to-MBL entanglement transition, on the other hand, are already fully interacting and their entanglement properties are thus expected to be robust to small local changes to the Hamiltonian.

\begin{figure}[tb]
\begin{center}
\includegraphics[width=0.49 \textwidth]{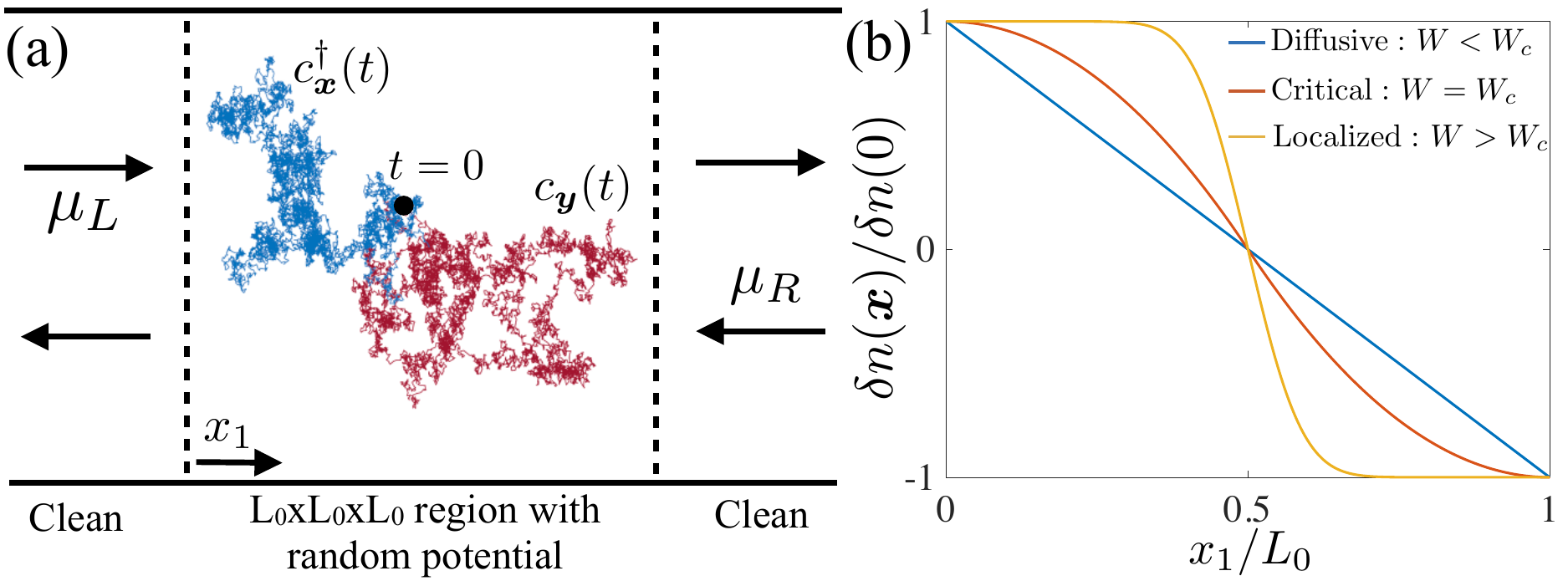}
\caption{(a) We study a noninteracting, boundary-driven fermionic system consisting of a cubic disordered region of size $L_0\times L_0\times L_0$ coupled to clean leads on both ends.  The left/right incoming scattering states (denoted by incoming arrows) are taken to be at thermodynamic equilibrium with the same temperature, but different chemical potentials $\mu_{L/R}$. Red and blue traces show the diffusive operator dynamics of an initially local density operator in the random circuit version of this model \cite{Gullans18}.  (b) Non-equilibrium density profile $\delta n(\bx)$ in the limit $L_0 \to \infty$ for the diffusive phase $0< W < W_c$, the critical point $W=W_c$, and the localized phase $W>W_c$ for $\xi/L_0 = 0.05$ and $a=0.25$.  In the localized phase, transport occurs  dominantly through a sub-extensive number of ``resonant'' states near the center of the sample.   
}
\label{fig:system}
\end{center}
\end{figure}

Much of our analysis applies quite broadly to any noninteracting model exhibiting an Anderson metal-insulator transition.  For concreteness, we focus on the setup shown in Fig.~\ref{fig:system}(a).  Two clean, semi-infinite quasi-1D wires with transverse dimensions $L_0 \times L_0$ are linked by a cubic disordered region of length $L_0$.  The Hamiltonian is given by 
\be \label{eqn:H}
H = \sum_{\mean{\bx \by}} c_\bx^\dag c_\by + \sum_\bx V_\bx c_\bx^\dag c_\bx,
\ee 
where $c_\bx$ are fermionic annihilation operators for site $\bx$, we work in units where the nearest-neighbor  hopping rate is one, and the quenched disorder at each site $V_\bx$ are drawn from independent uniform distributions between $\pm W/2$ ($V_\bx =0$ in the leads).  We assume periodic boundary conditions in the transverse directions, and use a simple cubic lattice.  The localization transition in the disordered region  occurs at a critical disorder strength in these units $W_c \approx 16.5$ \cite{MacKinnon81,Pichard81,Slevin14}.  We are interested in the non-equilibrium steady state (NESS) defined by the condition that the incoming scattering states from the left/right lead are in thermal equilibrium with the same temperature $T$ and different chemical potentials $\mu_{L/R}$.   More precisely, defining  $a_{nE}^\alpha$ as the fermionic annihilation operator for the incoming scattering states with energy $E$ in transverse channel $n$ and lead $\alpha$, we take
\begin{align}
\{ a_{nE}^\alpha,a_{mE'}^{\beta \dag} \} &=   \delta(E-E') \delta_{\alpha \beta} \delta_{m n},\\
\mean{a_{nE}^{\alpha \dag} a_{mE'}^\beta} &=  \delta(E-E') \delta_{\alpha \beta} \delta_{m n} n_E^\alpha,
\end{align}
where  $n_E^\alpha = [e^{(E-\mu_\alpha)/T}+1]^{-1}$ is the Fermi function.   It is further convenient to define sum and difference Fermi functions $n_{E}^{s,d}=(n_E^L \pm n_E^R)/2$.   To avoid complications  associated with bound states in the sample, we allow for leads with anisotropic hopping in the longitudinal ($x_1$) direction $t_{\parallel} > t_\perp $ \cite{Markos06}.  Similarly, to avoid mobility edge effects we take $\mu_{L/R}$ near zero energy with a chemical potential bias $\delta \mu = \abs{ \mu_L - \mu_R} \gg T$ and much less than the width of the mobility edge in the sample.

\emph{Mutual coherence.---}Due to the absence of interactions, all correlation functions of the NESS density matrix $\rho$ can be expressed in terms of the second-order correlation functions,
\be
G_{\bx \by} = \mean{c_\bx^\dag c_\by}= \trace[\rho \, c_\bx^\dag c_\by],
\ee
according to Wick's theorem \cite{Chung01,Cheong04,Peschel09}.  Particle conservation implies that $\mean{c_\bx c_\by} =0$.  The unique correspondence between the density matrix and the two-point function for Gaussian states motivates us to introduce the  ``mutual coherence'' as a particularly simple measure of entanglement and correlations between regions $A$ and $B$:
\be
C(A:B) =2 \sum_{\bx \in A, \by \in B} |\mean{c_\bx^\dag c_\by}|^2 + |\mean{c_\bx c_\by}|^2,
\ee
which measures the overall magnitude of spatial coherences between the fermions.  Within the set of  fermionic states, $C(A:B)$  serves as an entanglement witness because it is zero between all separable fermionic states.  Here, we define separable fermionic states with respect to a bipartition $A$ and $B$ as the set of states that can be formed by local fermionic operations and classical communication on $A$ and $B$ \cite{Banuls07}.  This definition implies that each region has a well defined fermionic parity so that the correlations
\begin{align}
\mean{c_i^\dag c_j} = \mean{c_i^\dag} \mean{c_j} = 0,~\mean{c_i c_j} = \mean{c_i} \mean{c_j} = 0,
\end{align}
vanish in a separable state for $i \in A$ and $j\in B$.  As a result, all such separable fermionic states have zero mutual coherence.
%To see intuitively why $C(A:B)$ is a natural measure of entanglement for Gaussian states, one can consider a many-body state at half-filling consisting of a product state of two-site fermion Bell pairs of the form $\frac{1}{\sqrt{2}}(\ket{0}_\bx \ket{1}_\by+ \ket{1}_\bx \ket{0}_\by)$.  Then $C(A:B)$ is directly proportional to the number of Bell pairs between regions $A$ and $B$.  More formally, within the set of Gaussian fermionic states, $C(A:B)$ is a proper measure of entanglement because it is zero between all separable Gaussian states \cite{Banuls07}, satisfies additivity, and is monotonically decreasing under Gaussian local operations and classical communication (LOCC).  
For Gaussian fermionic states, the mutual coherence is a lower bound on the mutual information \cite{supp}.  Moreover, for Gaussian states near infinite temperature, it accurately approximates both the mutual information and the fermionic entanglement negativity \cite{Shapourian18}.

We can move between the original fermionic operators and the scattering states using the scattering state wavefunctions $\phi_{nE}^\alpha(\bx)$ 
\be
c_\bx = \sum_{n \alpha} \int dE\, \phi_{nE}^\alpha(\bx) a_{nE}^\alpha,~ a_{nE}^\alpha = \sum_\bx \phi_{nE}^{\alpha *}(\bx) c_\bx,
\ee 
where the wavefunctions are normalized to have unit current in the incoming lead \cite{Pendry92}.  For a fermionic system whose incoming scattering states are at local equilibrium in each lead, the two-point function takes the form
\begin{align}
G_{\bx \by} &= G_{\bx \by}^s + G_{\bx \by}^d =  \int d E\, [ q_E^s(\bx,\by) +q_E^d(\bx,\by) ], \\ \nonumber
q_E^{s,d}&(\bx,\by)  = \sum_n [ \phi_{nE}^{L *}(\bx) \phi_{nE}^L (\by) \pm \phi_{nE}^{R *}(\bx) \phi_{nE}^R (\by) ]  n_{E}^{s,d},
\end{align}
where we have separated out the contributions to $G_{\bx \by}$ into an ``equilibrium'' ($s$) part that is symmetric under the exchange $\mu_L \leftrightarrow \mu_R$ and a ``non-equilibrium'' ($d$) part that vanishes when $\delta \mu=0$.  Time-reversal symmetry of $H$ implies that $G_{\bx \by}^s$ is real and carries zero current.

\emph{Diffusive phase.---}The non-equilibrium density profile across the transition is shown in Fig.~\ref{fig:system}(b).  In the diffusive phase, the coarse grained density profile follows from the steady-state solution to the diffusion equation $D \nabla^2 \delta n(\bx) = 0$: $\delta n(\bx)/\delta n(0) = 1 - 2 x_1/L_0$.  Here $D$ is the diffusion constant,  $\delta n(\bx)=\overline{ G_{\bx\bx}^d}$ is the non-equilibrium contribution to the density profile, and we have taken $\mu_L  = - \mu_R >0$.   

\begin{figure}[tb]
\begin{center}
\includegraphics[width=0.49 \textwidth]{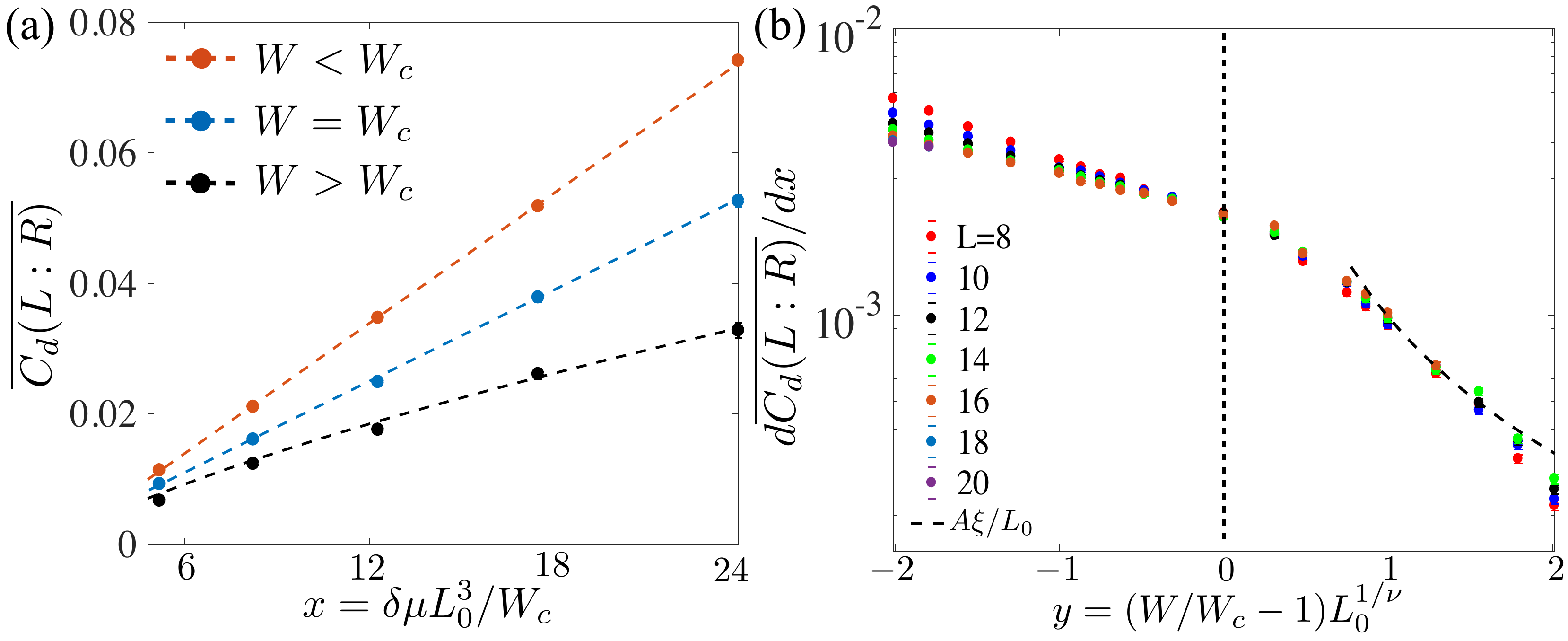}
\caption{(a) Scaling of $\overline{C_d(L:R)} $ between the left and right half of the sample in the diffusive phase ($W=10 $), the critical point ($W=16.5$), and in the localized phase ($W=21$).    We took a fixed chemical potential bias $\delta \mu/W_c = 3 \cdot 10^{-3}$ and varied  $L_0$ between 12 and 20.  The  red/blue/black dashed lines are fits to volume/volume/area-law scaling, respectively.   (b) Finite size scaling of $ d \overline{C_d(L:R)}/dx$. The derivative was evaluated at $x=24$ to ensure $\delta \mu/E_{\rm Th} \propto x /|y|^\nu > 1$ for $|y|>1$ and $\delta \mu$ is much larger than the level spacing near the critical point $|y| \ll1 $.  The black dashed line shows a fit to $A \xi/L_0$ on the insulating side, consistent with   a crossover to area law scaling for the mutual coherence. In both (a)-(b) we took $(t_\parallel,t_\perp) = (3 ,1)$ in the leads and $T=0$.   }
\label{fig:scaling}
\end{center}
\end{figure}

It was shown in our previous work that the mutual coherence (first defined here) exhibits a volume-law scaling in the diffusive phase \cite{Gullans18}.  An intuitive picture for this scaling was developed using a random circuit model, which can be realized in the present context by allowing both the nearest-neighbor hopping rates and disorder in $H$ to change randomly in time and space at discrete intervals.  The time-dependence of the parameters prevents localization and heats up the system, but with a density gradient between the left and right leads.  Evolving the coherences $|\mean{c_\bx^\dag c_\by}|^2$ under $H(t)$, one finds that they have an effective source term near $\bx = \by$  proportional to $\mean{\bm{J}(\bx)} \cdot \vec{\nabla} \mean{ n(\bx) }$, where $\bm{J}(\bx)$ is the current operator and $\vec{\nabla} n(\bx)$ is the local density gradient.  This can be interpreted as a microscopic realization of Ohm's law of dissipation.  A schematic picture of the subsequent operator dynamics for the coherences is shown in Fig.~\ref{fig:system}(a).  In effect, the coherences generated by the source live for a diffusive Thouless time $\tau_{\rm Th} = L_0^2/D$, before escaping into the reservoirs.  The time-averaged current density satisfies Fick's law $\overline{\mean{\bm{J}(\bx)}} = - D \vec{\nabla} \mean{n(\bx)}$, which leads to the scaling of the source term as $D [\delta n(0)]^2/L_0^2$.  Thus, the local production rate for the coherences scales as $\sim D/L_0^2$ and their lifetime scales as $\sim L_0^2/D$.   Defining the coherence density of site $\bx$ with a given region $A$ as $c_A(\bx)=C(A:\{\bx\})$, we can see that the coherence production rate balances with the decay rate to give an order one coherence density of a site in the bulk with the rest of the sample.  Crucially, these coherences are spread  fairly uniformly across the entire sample, which implies that this finite coherence density will persist when we take $A$ to be given by the left half the sample $L$.  Summing the coherence density over the right half of the sample $R$ gives rise to the volume-law scaling for $C(L:R)$.   To generalize this analysis to the time-independent case, one has to take into account the frequency dependence of the diffusion constant and other effects that arise due to energy conservation in this model.  We present a formalism in the supplemental material that allows one to include these effects in the scaling analysis.   Figure~\ref{fig:scaling}(a) presents numerical evidence for this volume-law scaling in the diffusive phase.   The non-equilibrium contribution to the mutual coherence $C_d(L:R) \equiv 2 \sum_{\bx \in L,\by \in R} |G^d_{\bx \by}|^2$ was computed from scattering state wavefunctions obtained via a transfer matrix method \cite{Markos06}.

\emph{Critical point.---}At the critical point ($W=W_c)$ in an infinite disordered system, single-parameter scaling theory predicts a scale dependent diffusion constant $D(\bx) \sim D_0/|\bx|$ \cite{Chalker90}.   In the case of the open geometry considered here, one can similarly describe the transport through the sample in terms of an inhomogeneous diffusion constant $D(\bx) = D_0/(1+x_{B}/x_c)$, where $x_{B} = \min(x_1,L_0-x_1)$ is the distance to the nearest boundary and $D_0$ and $x_c$ are free parameters \cite{Tiggelen90}.  The steady-state profile shown in Fig.~\ref{fig:system}(b) is  modeled with the solution to the diffusion equation $\vec{\nabla} \cdot D(\bx) \vec{\nabla} \delta n(\bx)  = 0$.  

In the case of the mutual coherence, we can find the local production rate for the coherences $\sim D(\bx)/L_0^2$ by applying similar arguments as in the diffusive phase.  The production rate in the bulk of the sample $\sim L_0^{-3}$ is suppressed by the scale-dependent diffusion constant.   However, the time for these coherences to reach the boundary  now scales as $\tau_{\rm Th} \sim L_0^3$.   Thus, we still expect an order one coherence density for each site in the bulk with the rest of the sample.  This coherence is again  spread fairly uniformly throughout the sample, leading to a volume-law scaling for $C(L:R)$.  Our numerical results shown in Fig.~\ref{fig:scaling}(a) agree with this scaling analysis.  Note that we take $\delta \mu$  much less than the width of the single-particle mobility edge, but still much greater than the single-particle level spacing in the sample $\sim L_0^{-3}$.  We leave a full analysis of the crossover at the mobility edge for future work.

\emph{Localized phase.---}For the localized phase, the physical mechanism underlying transport is quite distinct from the critical point and diffusive phase.  In this case, transport can only occur due to the exponentially weak overlap of the localized states  in the sample with both leads.  We refer to the localized states  near the center of the sample with nearly equal (but still exponentially small) tunneling rates to both leads as ``resonant'' states.  One signature that resonant states dominate transport is that the density profile exhibits a sharp step-like feature as shown in Fig.~\ref{fig:system}(b).  The width of the step is determined by the fluctuations in the tunneling rate of the resonant states to the leads, which directly maps to a well-studied problem in the statistics of directed paths in random media \cite{Huse85,Kardar85,Kardar94}.  In dimension $d$, one thus expects the width of the step to scale as  $\xi^{1-a} L_0^a$, where $\xi \sim |W - W_c|^{-\nu}$ is the localization length, $\nu \approx 1.57$ in 3D, and $a \approx 1/(d+1)$ \cite{Kardar94}.  One can partially account for these effects with a spatially varying diffusion constant of the form $D(\bx) \sim e^{-x_1(L_0-x_1)/\xi^{2(1-a)} L_0^{2 a}}$ \cite{Tian10,Tian13}, which was used to  model the density profile in  Fig.~\ref{fig:system}(b).  

In determining the mutual coherence, it is important to note that, although the current flowing through the  resonant states is exponentially small (leading to an exponentially weak production rate for the coherences), the slow production rate of coherences is  compensated by their exponentially long lifetime.  Thus, each point in the localized wavefunction of  a resonant state has order one coherence density with the rest of that state.  In the supplemental material, we provide an explicit calculation of this effect in a simplified 1D model for the resonant states as a two-mirror cavity \cite{supp}.   One  distinction from the diffusive phase and the critical point, however, is that these coherences are now confined within a localization length $\xi$ of the source due to the exponential localization of the wavefunctions.  As a result, we predict that the scaling for  $C(L:R)$ is  upper bounded by the area law $\sim \xi L_0^2$ in the localized phase  \cite{footnote1}.  Another important difference is that the spatial location of the resonant states fluctuates strongly within the sample on the macroscopic scale $\sim \xi^{1-a} L_0^a$.  This latter point implies that, deep in the localized phase $(\xi \ll L_0)$, the mutual coherence between the left and right half  has contributions from only a finite fraction of the resonant states $\gtrsim \xi^a/L_0^a$.  In single-parameter scaling theory, this could lead to the scaling for the mutual coherence with $\xi$ as $\xi^{1+b} L_0^{2-b}$ for some $0\le b <1$.    Our numerical results in Fig.~\ref{fig:scaling}(a)-(b) are consistent with an area-law scaling ($b=0$), but, due to the limited sizes we are able to access, we can not clearly resolve this point in the present work.

\emph{Scaling function.---}Assuming the validity of single-parameter scaling theory \cite{Abrahams79}, we can write a scaling function for the mutual coherence for $ \delta \mu $ much greater than $T$ and much less than the width of the mobility edge
\be
\overline{C_d(L:R)} = L_0^\alpha f[ \delta \mu L_0^3/W_c, (W/W_c - 1) L_0^{1/\nu}],
\ee
where the first argument $x=\delta \mu L_0^3/W_c$ measures $\delta \mu$ in units of the level spacing in the sample and the second argument is $y= (W/W_c - 1) L_0^{1/\nu} \propto (L_0/\xi)^{1/\nu}$.
According to our scaling analysis and numerical results at the critical point, the scaling dimension of the mutual coherence is $\alpha =0$.  Instead, the volume-law scaling arises  from the scaling function $f(x,y)$ being linear in $x$ at large values of $x$ for $y \leq 0$.  Figure \ref{fig:scaling}(b) shows our numerical finite size scaling analysis of $d\overline{C_d(L:R)}/d x$, where we see a collapse of the data for large systems sizes.  The numerical data is consistent with a crossover to area-law scaling in the localized phase based on the large $y$ behavior of the scaling function as  $f(x,y) \sim x/y^\nu \propto \delta \mu\, \xi L_0^2 $.

\emph{Conclusion.---}In this work, we  revisited the Anderson localization transition as an example of an entanglement phase transition in open quantum many-body systems.  
%We introduced the mutual coherence as a natural measure of entanglement and correlations for fermionic states.  We then determined the finite size scaling behavior of this measure across the localization transition.  The scaling of the mutual coherence at all points in the phase diagram can be derived by appealing to simple physical arguments based on the production, spreading, and decoherence of operators in this system.  
Future work could investigate the many-body localization transition from a similar perspective, where interactions may qualitatively change the scaling behavior on both sides of the localization transition.  Another promising direction is to  experimentally study the mutual coherence in driven-disordered systems accessible by local probes such as ultracold atoms, two-dimensional condensed matter systems, or scalable quantum information platforms.   

\begin{acknowledgments}
We thank Sarang Gopalakrishnan for helpful discussions.  Research supported in part by the DARPA DRINQS program,  DARPA grant No. D18AC0025,  and the Gordon and Betty Moore Foundation's EPiQS Initiative through Grant GBMF4535.  
\end{acknowledgments}

%\bibliographystyle{apsrev-nourl-title-PRX}
%\bibliography{Chaos}

\bibliography{EntTrans_Lett_v4.bbl}

\begin{thebibliography}{77}
\bibitem[{foo()}]{footnote1}
\bibinfo{note}{Note that these arguments do not depend on the precise details
  of the envelope wavefunctions (e.g., exponentially decaying versus a power
  law), so long as the lifetime of the resonant states is much less than the
  noninteracting level spacing in the sample, justifying a perturbative
  treatment of their coupling to the leads.}
\end{thebibliography}

\pagestyle{empty}
{ 
\begin{figure*}
\vspace{-1.8cm}
\hspace*{-2cm} 
\includegraphics[page=1]{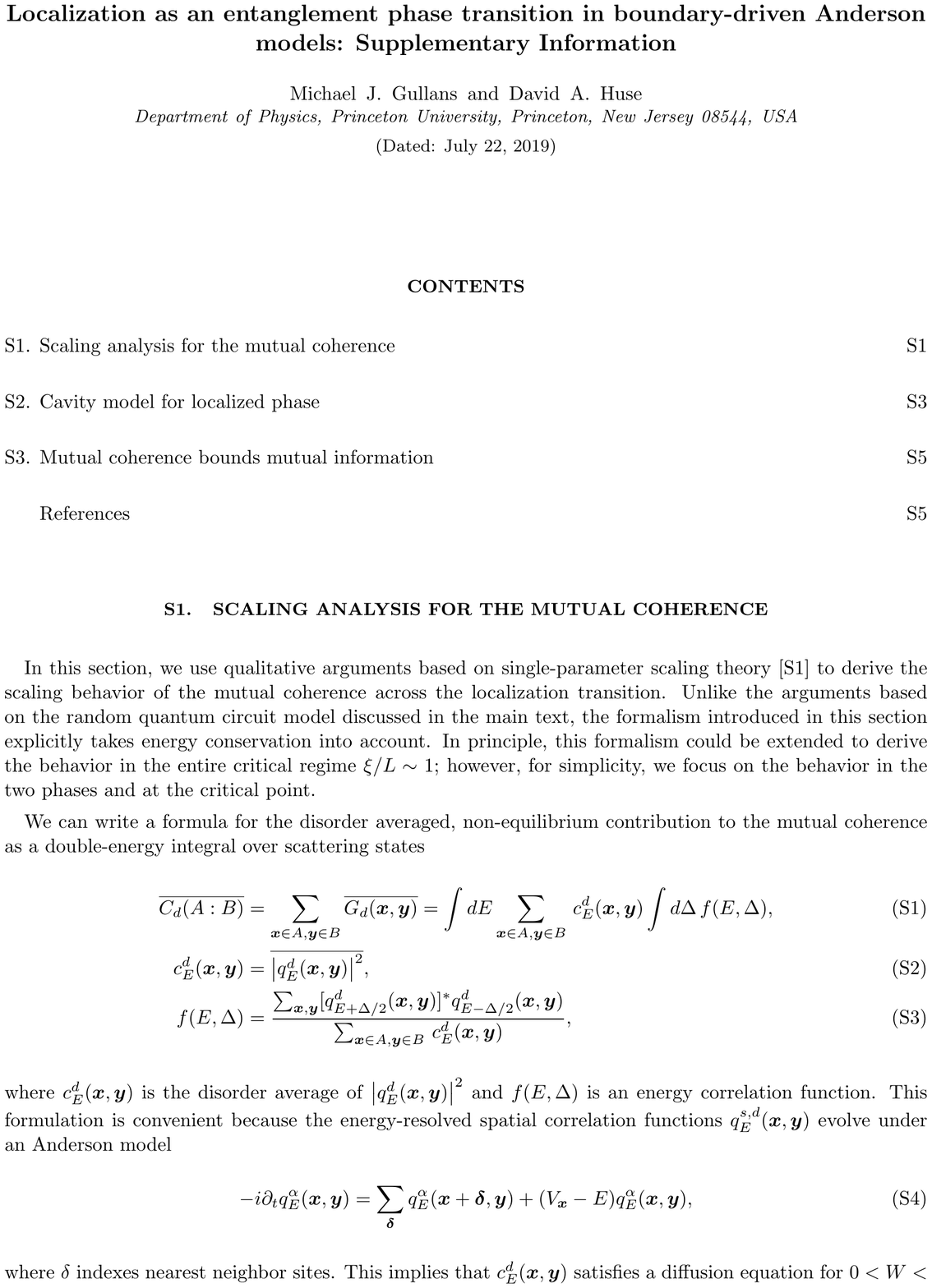}
\end{figure*}

\begin{figure*}
\vspace{-1.8cm}
\hspace*{-2cm} 
\includegraphics[page=2]{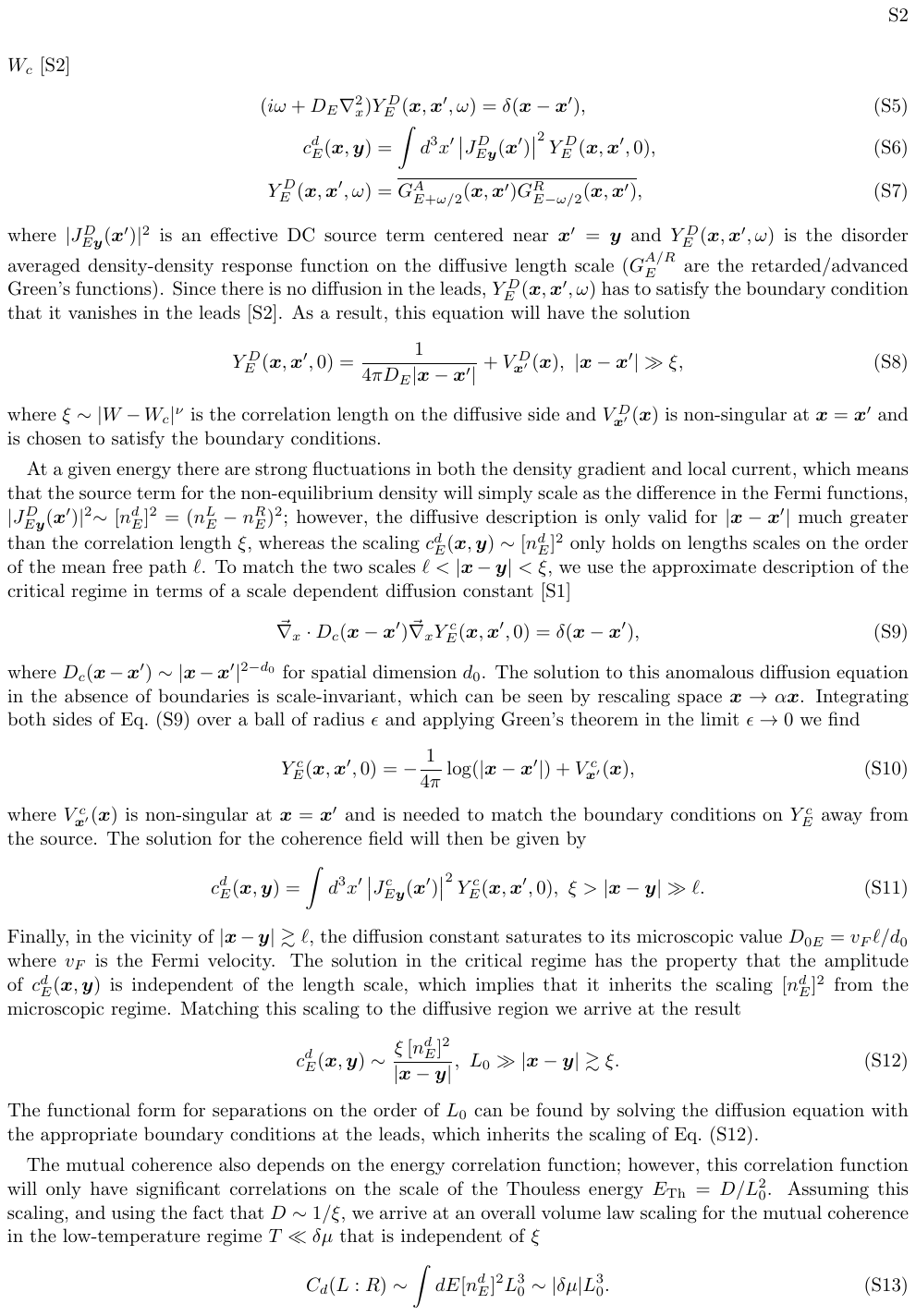}
\end{figure*}

\begin{figure*}
\vspace{-1.8cm}
\hspace*{-2cm} 
\includegraphics[page=3]{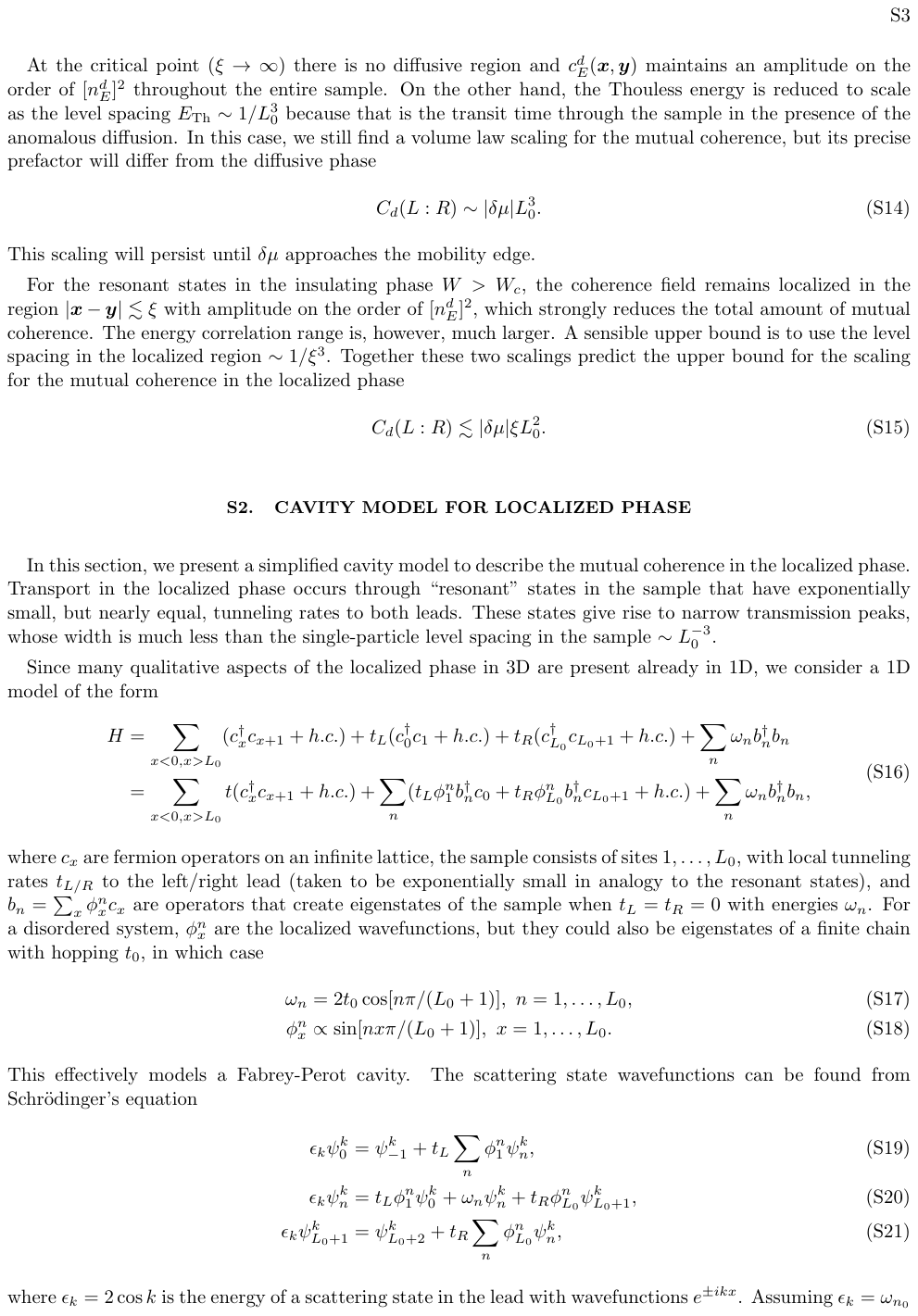}
\end{figure*}

\begin{figure*}
\vspace{-1.8cm}
\hspace*{-2cm} 
\includegraphics[page=4]{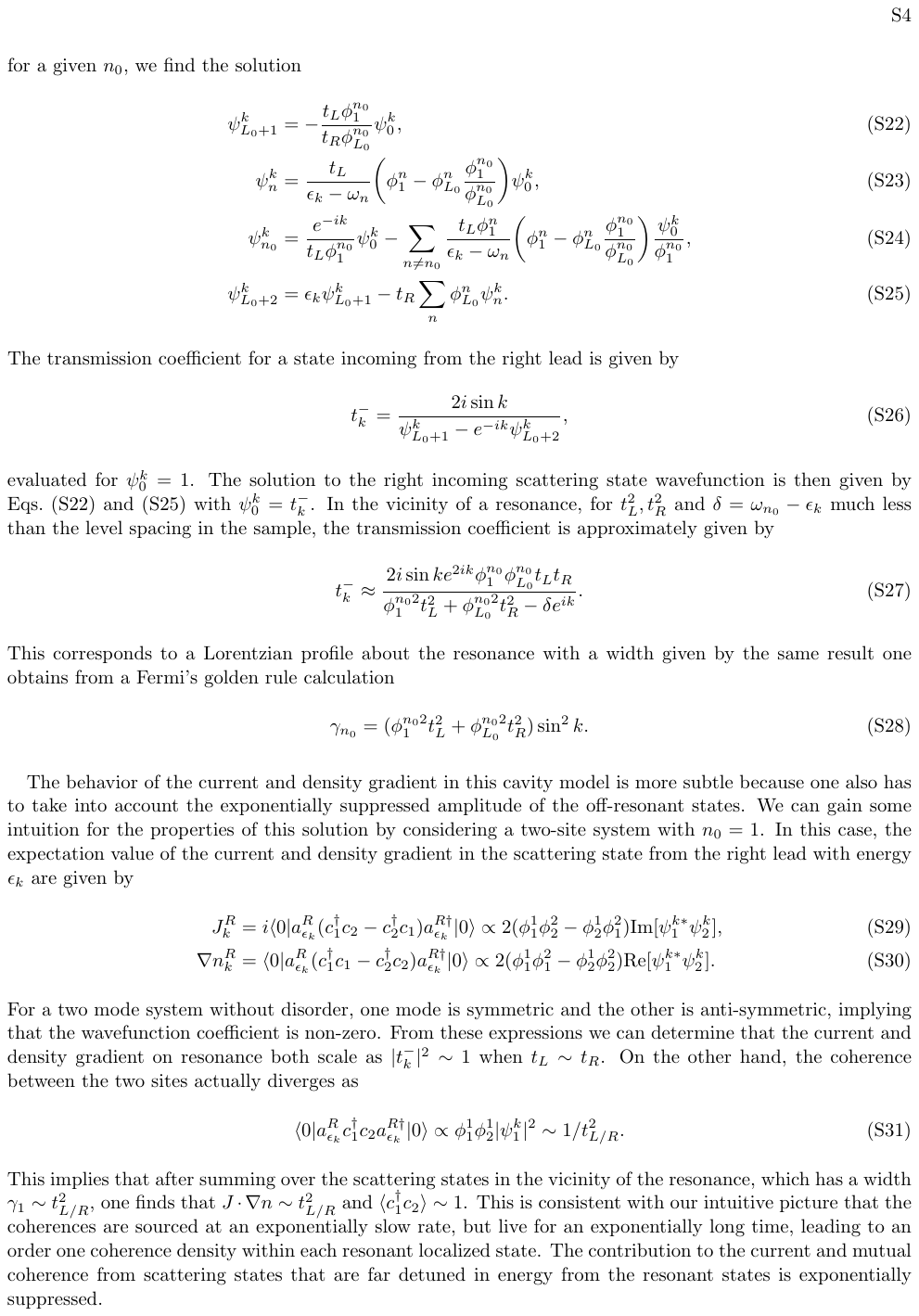}
\end{figure*}

\begin{figure*}
\vspace{-1.8cm}
\hspace*{-2cm} 
\includegraphics[page=5]{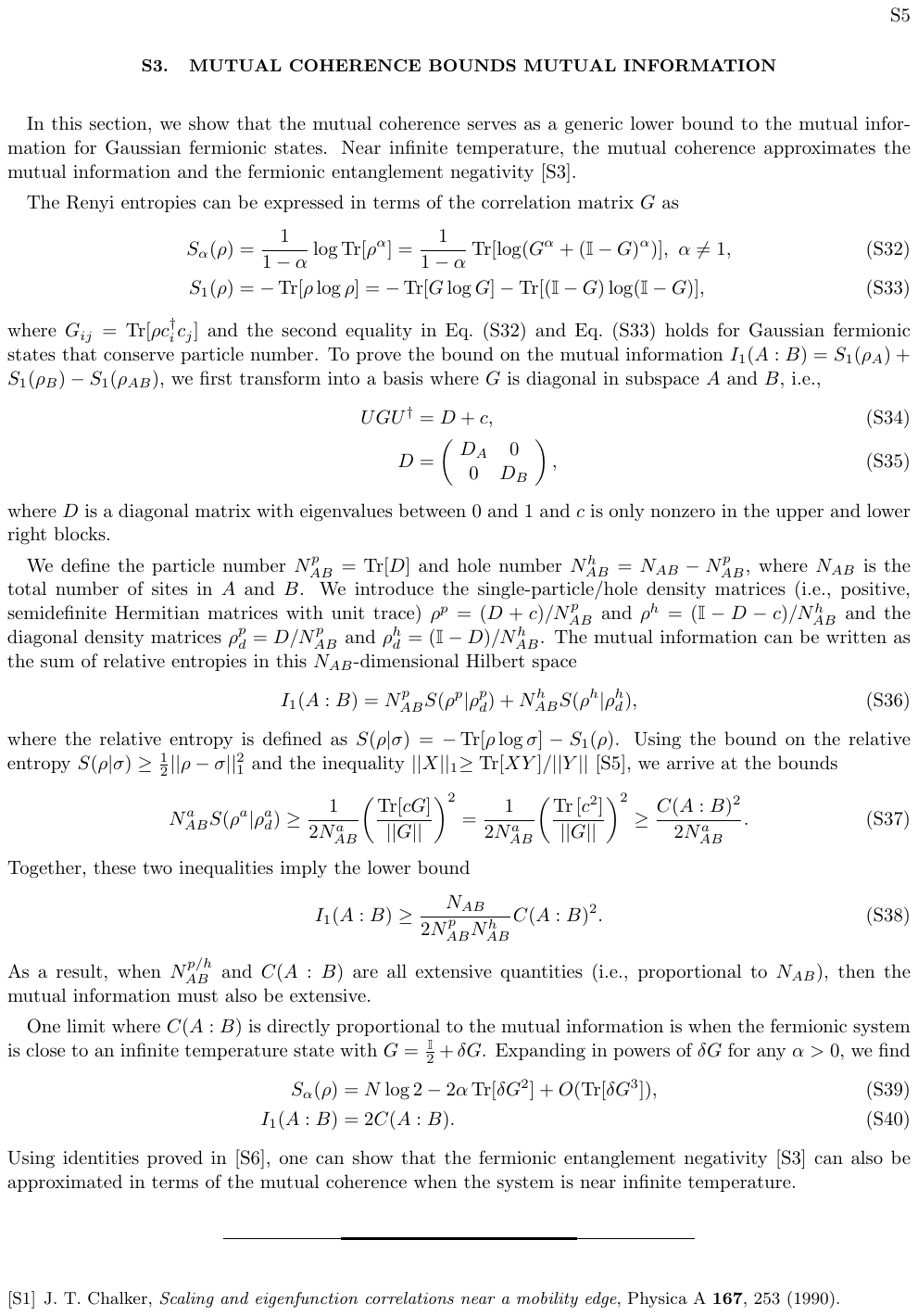}
\end{figure*}

\begin{figure*}
\vspace{-1.8cm}
\hspace*{-2cm} 
\includegraphics[page=6]{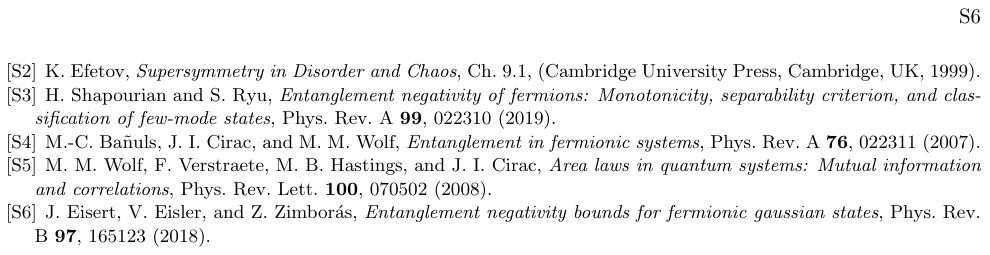}
\end{figure*}
}

\end{document}